# A New Method for Searching for Free Fractional Charge Particles in Bulk Matter[*]

Dinesh Loomba, Valerie Halyo, Eric R. Lee,
Irwin T. Lee, Peter C. Kim, and Martin L. Perl

*Stanford Linear Accelerator Center*
*Stanford University, Stanford, CA 94309*

ABSTRACT

We present a new experimental method for searching for free fractional charge in bulk matter; this new method derives from the traditional Millikan liquid drop method, but allows the use of much larger drops, 20 to 100 µm in diameter, compared to the traditional method that uses drops less than 15 µm in diameter. These larger drops provide the substantial advantage that it is then much easier to consistently generate drops containing liquid suspensions of powdered meteorites and other special minerals. These materials are of great importance in bulk searches for fractional charge particles that may have been produced in the early universe.

*Submitted to* Review of Scientific Instruments

---

[*] Work supported by Department of Energy contract DE-AC03-76SF00515.

1. INTRODUCTION

The Millikan liquid drop method is the most robust of the various methods for searching for fractional charge particles [1-3]. The traditional Millikan method, Section 2, as described by Perl and Lee [4] and by Mar *et al.* [5], can be used to investigate a broad range of charges with a resolution that currently exceeds 1/40 of an electron's charge [5,6]. In addition the mass search range for the fractional charge particle is many orders of magnitude larger than the mass search range of existing or planned accelerators. Furthermore, unlike levitometer experiments [1-3], that have been plagued by a lack of charge calibration, the Millikan method achieves large statistics that provide a natural self-calibration of the charge. Finally, the method is inherently amenable to automation and is easily replicable.

In the past the traditional Millikan method was only used with liquids such as mercury, sea water, and silicone oil [5-8]. Hendricks *et al.* [9] proposed an extension to the pure liquid Millikan method that allowed the examination of arbitrary materials, solids as well as liquids. In their approach solid search materials are pulverized and incorporated into colloidal suspensions, Appendix A. Drops can then be formed from these suspensions in the same manner as for liquids, thus extending the Millikan technique beyond liquid search materials.

As qualitatively discussed in Appendix A, the *consistent* generation of drops containing suspensions of pulverized materials is much easier in larger drops. The purpose of the new method, Section 4, is to allow the use of larger drops, above 20 μm in diameter, compared to the less than 15μm diameter drops that we use in the traditional Millikan method [4-6].

Before proceeding to describe the new method it is useful to remind the reader of the importance of using meteoritic materials and other special materials when searching for fractional charge particles that may have been produced in the early universe [4]. Most levitometer and Millikan method searches for fractional charge particles have been carried out in *refined* bulk matter such as iron and niobium. As discussed by Lackner and Zweig [10] fractional charge particles usually change the chemical properties of the atoms containing the particles, sometimes in ways difficult to predict. Therefore when considering upper limits on the abundances of fractional charge particles in the solar system, it is not possible to deduce with certainty those limits from searches for those particles in *refined* materials. Therefore some of the most interesting materials are meteorites or moon rocks that are studied *without* being chemically processed when preparing the suspensions. Chondritic meteorites are especially attractive because they are generally believed to be representative of the earliest solar system matter in its most pristine state [11]. Other interesting materials are certain terrestrial minerals that may more readily accumulate elements containing fractional charge particles [4].

2.  **THE TRADITIONAL MILLIKAN METHOD**

The Millikan method, referred to in this paper as the traditional method, is schematically shown on Fig. 1a. Drops with diameters in the range of 6 to 12 µm are produced using a piezoelectric drop generator of our own design, Fig. 1b [4,5,12]. The exact diameter depends on how we set the parameters of the generator: aperture size, aperture shape, pulse shape, and pulse frequency. Once these generator parameters are set, the drop diameter is maintained to about 0.1%.

The drops fall through air, the forces on the drops are that due to gravity, that due to the resistance of the air, and that due to a vertical electric field; the field points alternately up and down. When the field points down the drop terminal velocity is $v_{down}$, when the field points up the terminal velocity is $v_{up}$. The terminal velocities are determined by measuring the vertical position, $z$, of the drop at time intervals of $\Delta t$. In our use of the traditional Millikan method we make of the order of ten to twenty successive determinations of the vertical position, $z_1, z_2, z_3…$; this increases the precision of our knowledge of the terminal velocities and checks that the drop charge has not changed during the measurement period.

The number of successive $z$ measurements is limited by the total number of pixels in the sensitive area of the CCD video camera in Fig. 1a, in particular the number of successive $z$ measurements is set by the pixel resolution in the $z$ direction. For example,

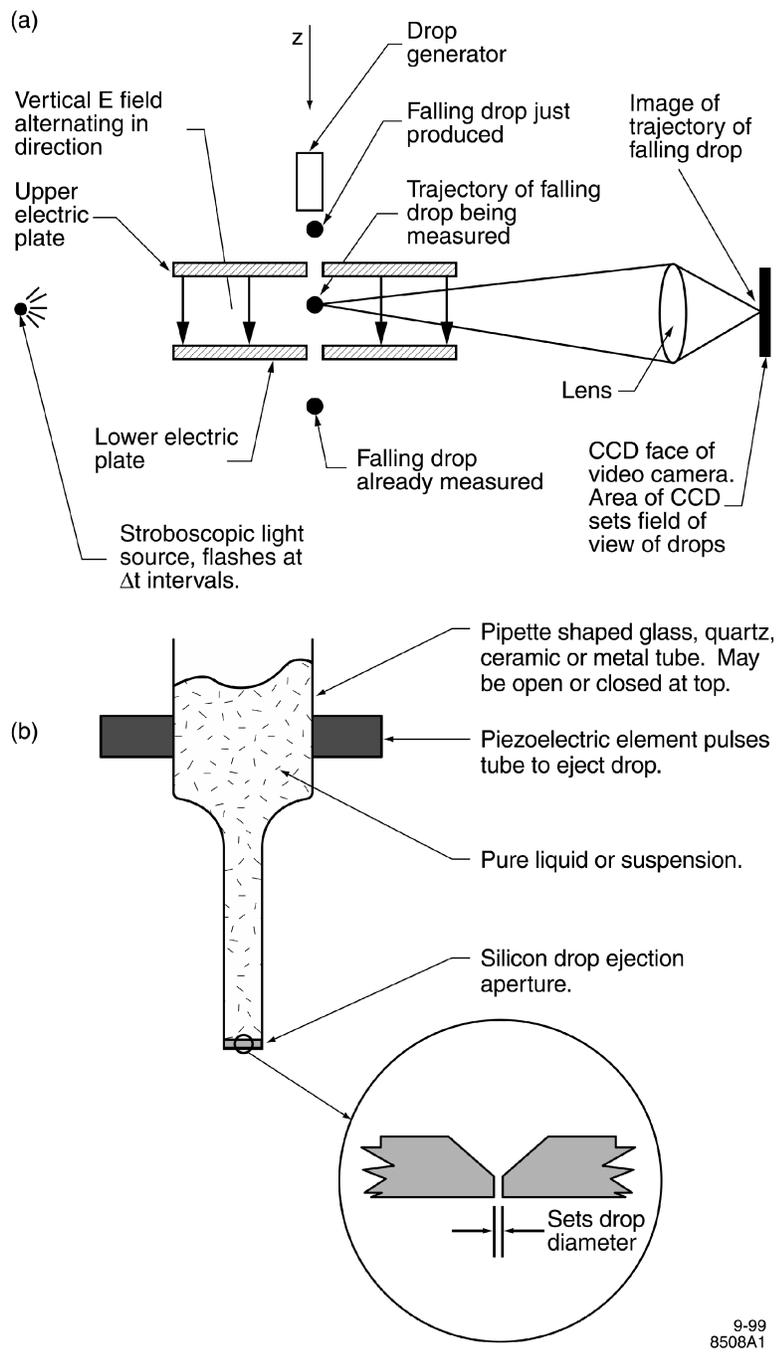

Fig. 1. (a) Schematic vertical view of our traditional Millikan method apparatus. (b) Explanatory view of the piezoelectric drop generator. The diameter of the hole in the drop ejection aperture plate sets the diameter of the drop within a range of about ± 15%. The exact diameter depends upon aperture shape, pulse shape, and pulse frequency

we use a CCD camera with 750 lines in the $z$ direction. One can increase the number of $z$ measurements by reducing the magnification from the drop space to the CCD face. But if the magnification is too small, the CCD provides insufficient $z$ resolution and this limits the precision of the charge measurement rather than the effect of Brownian motion discussed below.

The terminal velocities are given by the equations:

$$\frac{4}{3}\pi r^3 \rho g + qE = 6\pi \eta r v_{down} \tag{1}$$

$$\frac{4}{3}\pi r^3 \rho g - qE = 6\pi \eta r v_{up}.$$

Here, $r$ is the drop radius, $\rho$ is the mass density of the drop less the mass density of air, $q$ is its charge, $E$ is the magnitude of the electric field (taken to be the same for both directions), and $\eta$ is the viscosity of air.

There are two ways to determine $r$. First, $r$ is given by these equations if we know $\rho$, as we always do for pure liquids. In the case of suspensions there is a small uncertainty in the value of $\rho$ from drop to drop; then we use the second method of measuring $r$ directly from the drop image on the CCD. Knowing $r$, it's simplest to think of the drop charge as being given by

$$q = \frac{3\pi \eta r}{E}(v_{down} - v_{up}). \tag{2}$$

Ignoring any errors from the measurement of the $z_i$'s, and errors in the determination of $r$, the lower bound on the error in $q$ is due to the Brownian motion of the drops, which gives an error in the velocity measurement,

$$\sigma_v = \sqrt{\frac{kT}{3\pi \eta r \Delta t}}. \tag{3}$$

Here, $k$ is Boltzman's constant, $T$ is the air temperature, and $\Delta t$ is the time duration between successive measurements of the drop's position. We have found that this lower limit can almost be achieved (*i.e.,* that there is no other major measurement error), in which case the propagated error on the charge is given by [5]:

$$\sigma_q = \frac{3\sqrt{2}\pi\eta r \sigma_v}{E} = \frac{\sqrt{2}}{E}\sqrt{\frac{3\pi\eta r kT}{\Delta t}} \; . \tag{4}$$

In our experiments we reduce the error in measuring the charge by making $N_{pair}$ pairs of independent measurements of $v_{up}$ and $v_{down}$, then

$$\sigma_q(N_{pair}) = \sigma_q / \sqrt{N_{pair}} \; . \tag{5}$$

We use $N_{pair}$ in the range of 5 to 8. With drop diameters in the range of 6 to 12 µm as previously noted, we measure $q$ with a $\sigma_q(N_{pair})$ of .015$e$ to .025$e$, where $e$ is the electron charge. These values of $\sigma_q(N_{pair})$ are what we need for our fractional charge particle searches [4-6].

## 3. THE LARGE DROP PROBLEM IN THE TRADITIONAL MILLIKAN METHOD

As discussed in Appendix A, it is much easier to consistently generate drops containing suspensions of pulverized materials when the drop diameters are larger, above 20µm. Let us see what happens to $\sigma_q(N_{pair})$ when we increase the drop radius, $r$. The discussion in Section 2 shows that for fixed $N_{pair}$, $E$, and $\Delta t$, $\sigma_q(N_{pair})$ increases as $r^{1/2}$. But when we try to increase $N_{pair}$ in a practical apparatus so as to reduce $\sigma_q(N_{pair})$, Eq. 5, we face a double problem. First, $N_{pair}$ is limited by the maximum $z$ length between the electric plates viewed by the optical system, Fig. 1a. Eventually the drop falls out of the field of view of the optical system. Second, the average terminal velocity from Eq. 1 is

$$v_{ave} = \frac{2r^2 \rho g}{9\eta}.  \tag{6}$$

Therefore, increasing the drop radius $r$ increases $v_{ave}$ as $r^2$, and for fixed $\Delta t$ there are less measurements of $z$ and hence of $N_{pair}$. Finally as shown in Eq. 4, if $\Delta t$ is reduced $\sigma_q$ will increase.

In principle the problem of using larger drops in the traditional Millikan method apparatus can be solved by using an optical system and CCD camera with a much larger number of pixels along the z direction and by increasing the distance between the electric plates in Fig.1a. In practice this solution has many problems, for example, such a fast, affordable CCD camera does not exist currently. Another problem is that increasing the distance between the electric plates increases the required voltage difference between the plates. Switching the voltage to give alternate up and down electric fields then becomes more difficult.

## 4. THE NEW METHOD

In the new method, shown in Fig 2, the electric field direction is horizontal and thus orthogonal to the direction of the gravitational force. In this configuration the motion in the vertical direction, $z$, depends on $r$ and $\rho$, and the motion in the horizontal direction, $x$, depends on $r$ and $q$:

$$v_x = qE/6\pi\eta r \tag{7a}$$
$$v_z = 2r^2\rho g/9\eta \tag{7b}$$

Thus the effect of gravity on the drop trajectory is decoupled from that of the electric field and the charge is computed solely by measurement of $v_x$,

$$q = 6\pi\eta r v_x / E.  \tag{8}$$

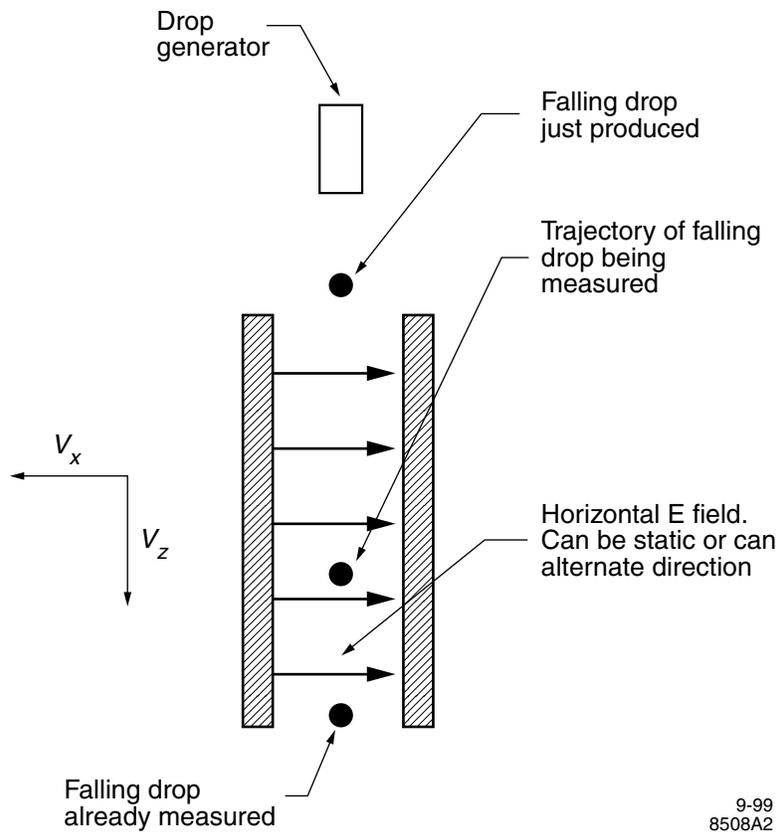

Fig. 2. Schematic illustration of the use of a horizontal electric field. The view is along the axis of the optical system.

As before, we assume that $r$ is measured independently and perfectly, and that the Brownian motion of the drop dominates the error in $q$, then

$$\sigma_q = \frac{2}{E}\sqrt{\frac{3\pi\eta r kT}{\Delta t}}. \qquad (9)$$

This is $\sqrt{2}$ times worse than that of Eq. 4 as expected, since the latter requires two independent measurements of velocity to extract $q$ whereas here only one is needed.

Therefore, insofar as charge resolution is concerned, the new method is equivalent to the traditional method for the same number of independent velocity measurements.

In this new method, the electric field can be static or it can be alternated in direction as in the traditional Millikan method. Either choice has advantages. With a static field, the drops move smoothly along straight-line trajectories prescribed by their charge. Trajectories corresponding to differently charged drops diverge linearly with time, which, if exploited, provides a means for improving charge resolution. With an alternating field, the drop trajectory remains within a narrow range of $x$, simplifying corrections.

In either case the greatest advantage of the new method originates from the independence of charge determination from measurement of the $z$ component of the drop's trajectory: $z$ component measurements can be ignored or altered, if needed to aid the experiment. This allows increasing the drop size while maintaining good charge resolution. We present next three ways to achieve this desirable experimental situation.

### A. Use of upward airflow

Recall that the worsening of charge resolution that accompanies an increase in drop size can be compensated *if* a corresponding increase in the number, $N_v$, of $v_x$ velocity measurements could be made. Recall, on the other hand, that in the traditional method with a fixed $z$ observation length, a decrease in $N_{pair}$ occurs because $v_z$ varies as $r^2$. However in the new method $v_z$ has nothing to do with the charge measurements, hence we are free to alter $v_z$ by introducing a laminar, upward directed airflow against the direction of the gravitational force, Fig. 3. Thus the drops can be slowed vertically as needed to achieve the necessary number of $v_x$ measurements.

Consider the following example, a 10 $\mu$m diameter drop falls with a $v_z$ of 3.0 mm/s. Suppose the vertical extent of the field of view of the optical system allows $N_{v,10}$ measurements of $v_x$. Here the subscript 10 denotes the diameter of the drop. Then with $\sigma_{q,10}$ given by Eq. 9, the final $\sigma_{q,10}(N_{v,10}) = \sigma_q/\sqrt{N_{v,10}}$. Now consider a 20 $\mu$m diameter drop. From Eq. 9, $\sigma_{q,20} = \sqrt{2}\,\sigma_{q,10}$. Also $v_z$ is now 12.0 mm/s which would reduce $N_v$ by a factor of 4. But suppose the upward airflow is 10.5 mm/s so that the $v_z$ of the 20 $\mu$m drop seen by the optical system is 1.5 mm/s, half the $v_z$ of the 10 $\mu$m drop. This give $N_{v,20} = 2 N_{v,10}$ and

$$\sigma_{q,20}(N_{v,20}) = \frac{\sigma_{q,20}}{\sqrt{N_{v,20}}} = \frac{\sqrt{2}\sigma_{q,10}}{\sqrt{2N_{v,10}}} = \frac{\sigma_{q,10}}{\sqrt{N_{v,10}}} = \sigma_{q,10}(N_{v,10}). \tag{12}$$

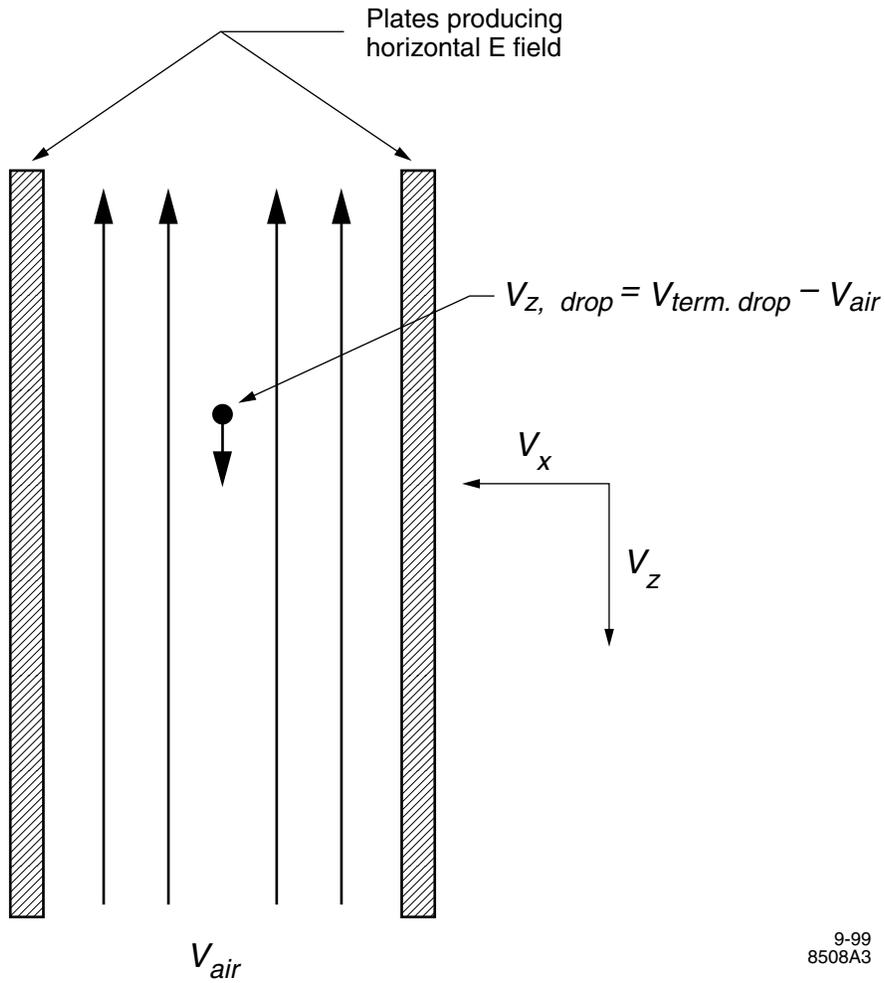

Fig. 3. Schematic illustration of the use of upward airflow to reduce drastically the downward vertical velocity, $v_{z,drop}$, of the drop. The view is along the axis of the optical system.

Laminar air flows with velocities of the order of several cm/s are easily achieved.

Although a near perfect laminar flow can be created at the low air speeds needed, a number of questions remain to be answered. The no-slip boundary condition at the walls of the chamber is responsible for producing velocity gradients in the airflow (Poiseuille flow). It is not known for certain whether small drops will suffer transverse forces from this effect. If they do, and these are uncorrectable, a plug-flow may be needed [13].

### B. Use of anamorphic optics

In our present search apparatus and in the method described in the previous section, the optical system has the same magnification in the horizontal and vertical planes. However, it is possible to use smaller vertical magnification compared to the horizontal magnification, then the optical system provides a larger field of view in the vertical direction, Fig. 4. Thus, instead of altering the $z$ trajectory of the drop, this method alters the *viewing* of the component, by compressing the magnification in that direction to increase $N$, resulting in images of pancake shaped drops. If successful, this method may be preferable to altering actual trajectories, as required for the airflow technique. The main issue that must be addressed is whether drop centroids and radii can be extracted from the distorted images. For maximal distortion the drop image reduces to a line (one pixel high in the $z$ direction) and centroiding might be possible from the intensity distribution along the line. The drop radius, in principle, could be extracted from the total intensity of the line. Since $v_x$ decreases with radius $\left(v_x \propto 1/\sqrt{r}\right)$ the magnification along $x$ can be increased to give a higher pixel density along this direction, which would aid in the measurement of both quantities.

### C. Use of long time interval $\Delta t_{long}$

In the traditional method and in the methods discussed in Sections 4A and 4B, numerous small time intervals, $\Delta t$, are used to make $N$ measurements of $v_x$, all equal in precision. In this method, for use in a static $E$ field, one long time interval, $\Delta t_{long}$, is used to make one very precise measurement of $v_x$. As shown in Fig. 5, the distance $\Delta z_{long}$ between the first and second $z$ measurement is too long to be viewed by a single optical-CCD system. Therefore, two optical-CCD systems are used.

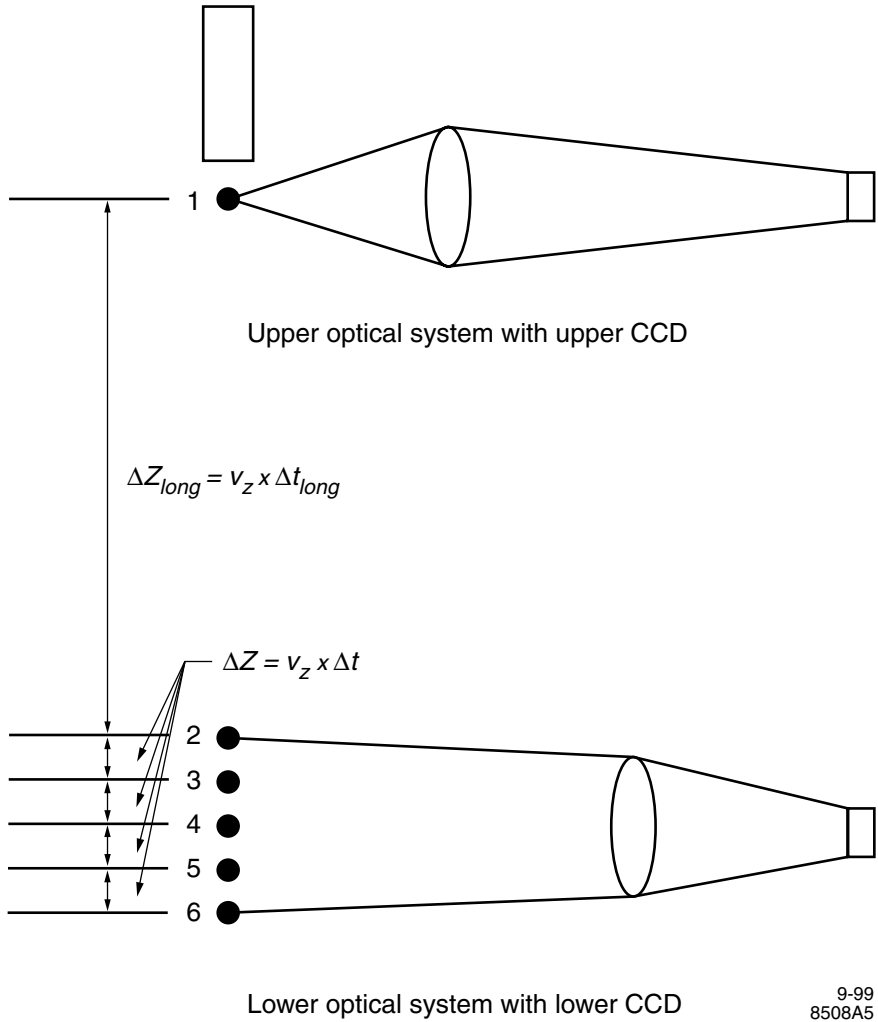

Fig. 4. Schematic illustration of the use of an anamorphic optical system with a CCD video camera. In the horizontal plane the object plane of falling drops (a) is magnified by 2 to make full use of the resolution of the image plane on the CCD face (b), thus determining $v_x$ with improved precision. In the vertical plane poorer resolution is acceptable, hence the magnification is 0.5, giving elliptical drop images. This allows more measurements of $v_x$.

The accuracy of this technique is derived from the large $\Delta z_{long}$ between the first two measured positions of the drop. Although the remaining $z$ position measurements by the lower optical system do not add much weight to the charge resolution; they are crucial for eliminating possible backgrounds because they provide redundancy. One such background is the possible occurrence of a charge change by the drop in its flight through the air, which, if it occurs between the first two position measurements, would result in a false signal for fractional charge.

## 5. REMARKS

The status of our work on the three methods described in Section 4 is as follows. We have built a first apparatus with a horizontal electric field and an upward airflow as described in Section 4A. Preliminary studies with 20 $\mu$m diameter drops show the method to be promising.

The method in Section 4B using anamorphic optics is now being studied, particularly with respect to optical system aberrations that might increase $\sigma_q$. The method using a long time interval, Section 4C, is not being considered for use in the near future.

When we first began to consider the use of larger drops for suspensions of pulverized materials, we also thought that there would be a substantial increase in the rate at which material could be studied, perhaps in proportion to the cube of the drop radius. However there are a number of rate questions that are still being studied. For example, falling drops exert aerodynamic forces on each other and we correct the charge measurements for this effect to first order [6]. But the effect is greater for larger drops.

## 6. ACKNOWLEDGEMENTS

We are grateful to Richard Blankenbecler for conversations about anamorphic optics. This work was supported by the Department of Energy contract DE-AC03-76SF00515.

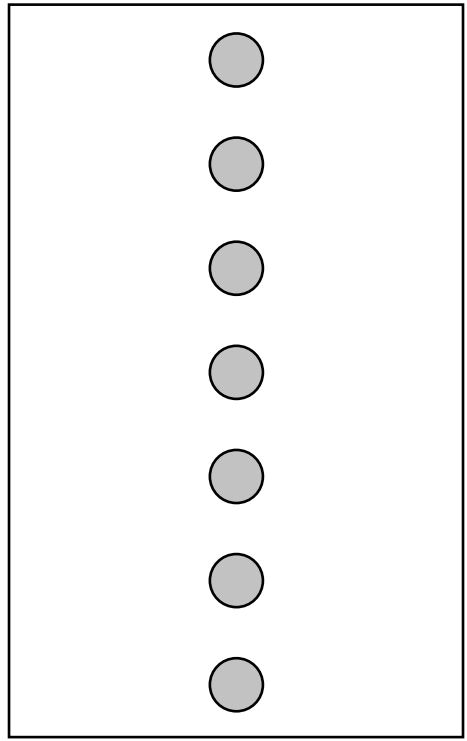

(a) Object plane of falling drops

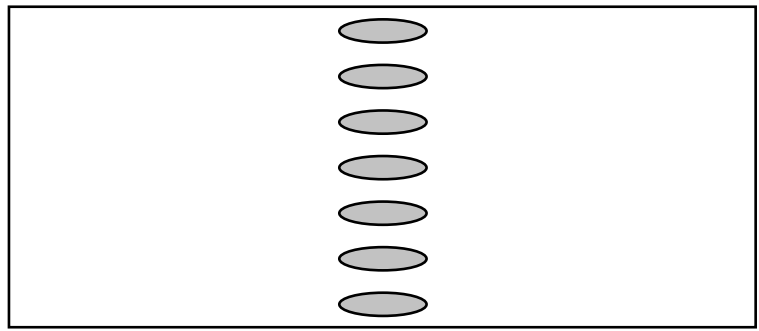

10-99
8508A6

(b) Image plane on CCD camera face

Fig. 5. Schematic vertical view of the use of a large $\Delta t_{long}$ to obtain a precise measurement of $v_x$ and hence of the drop charge. The precision is obtained between drop positions 1 and 2. The $v_x$ measurements made between positions 2 and 3, 3 and 4, and so forth, are used to confirm that the drop has not changed charge during the passage from positions 1 to 2. The upper optical system with its large magnification is also used to make a precise measurement of the drop radius $r$.

**APPENDIX A**

The purpose of this appendix is to explain the importance of using relatively large drops when searching in suspensions of meteoritic and other mineral materials for fractional charge elementary particle. In preparing these suspensions we mechanically pulverize the material into a powder fine enough to be suspended, we do *not* chemically process the material in any way. For example, we do not dissolve the minerals in acid and then attempt to form a fine precipitate.  The reason for our ban on chemical processing is that we have no way of knowing whether or not we might lose a fractional charge particle during the processing. There are several loss mechanisms in this type of process: the charged particle might be collected by the walls of the processing vessels; the charged particle, with or without its associated atoms, might not go into solution; or the charged particle, with or without its associated atoms, might not precipitate.  As discussed by Lackner and Zweig [10,14], the chemical behavior of atoms associated with fractional charge is complicated and sometimes difficult to predict.

In the following brief discussion of pulverization technology, recall that up to the present we have used drops with diameters less than 15 μm in diameter, and in this paper we are presenting methods for using drops of greater than 20 μm diameter. Our first step in pulverizing mineral samples is to use a ball mill or a mortar and pestle. A rule of thumb in pulverization technology is that such methods can only reduce many minerals to small pieces whose minimum size is of the order of ten or several tens of μms. There are two reasons, first minerals break at defects such as grain boundaries, and the smaller the piece the less the chance it contains defects. Second, as the mineral pieces get smaller they protect each other against further milling.

Our second step in pulverization is to use a jet pulverizer, a device in which high-speed air jets entrain the powder from the first step, the mineral pieces being given relatively high kinetic energies. These pieces then collide with each other further reducing their sizes.  In minerals with which we work, this produces a powder with sizes ranging from about 0.1 μm to about 5 μm. It is our experience and that of others, that further passes through the jet pulverizer *do not* substantially change the size distribution.

We find that when we generate drops of less than 12 to 15 μm diameter, containing a suspension of powdered mineral with powder sizes in the 0.1 μm to 5 μm range, the drop generator, Fig. 1b, works poorly. Sometimes it stops working completely; sometimes the mineral powder remains in the ejector tip with the generated drops containing little or no mineral powder.

Of course we could use just the fine end of the 0.1 μm to 5 μm powder to make a suspension, using say just 0.1 μm to 1 μm powder. But this would lead to a biased search of the mineral for fractional charge particles. Recall that the minerals we use, such as meteorites, are a complicated mixture of many different compounds such as silicates and oxides of different elements. As an extreme example, suppose that a particular compound in the mineral is the best vehicle for entraining fractional charge particles, and suppose that compound is the hardest to pulverize.

Our solution to this experimental dilemma is obvious, we use larger drops, about 20 μm or larger in diameter. Then with mineral powders in the size range of 0.1 μm to 5 μm we can get consistent generation of drops containing suspensions of these minerals.